\documentclass[prd,aps,preprint,epsfig,floats]{revtex4}
\input epsf.tex
\def\DESepsf(#1 width #2){\epsfxsize=#2 \epsfbox{#1}}
\begin{document}

%\draft

%@twocolumnfalse\endcsname
\preprint{\vbox{
\hbox{UMD-PP-03-047}
}}

\title{{\Large\bf Lepton Flavor Violation and Neutrino Mixings in a
$3\times 2$ Seesaw Model}}
\author{\bf Bhaskar Dutta$^1$ and R.N. Mohapatra$^2$ }

\affiliation{$^1$ Department of Physics, University of Regina, Regina, SK, S4S 0A2,
Canada\\$^2$ Department of Physics, University of Maryland, College Park,
MD 20742, USA}
\date{May, 2003}

\begin{abstract}
We study the predictions for $BR(\mu\rightarrow e+\gamma)$ and
$BR(\tau\rightarrow e,\mu+\gamma)$ in a class of recently proposed
horizontal $SU(2)$ model that leads to a $3\times 2$ seesaw model for
neutrino masses. We consider two such models and obtain the
correct bi-large mixing pattern for
neutrinos. In these models, the effective low energy theory below the
$SU(2)_H$ scale  is the MSSM.
Assuming a supersymmetry breaking pattern as in the minimal SUGRA models,
we find that consistent with present g-2, $b\rightarrow s+\gamma$ and WMAP
dark matter constraints on the
parameters of the model, the $BR(\mu\rightarrow e+\gamma)$ prediction is
accessible to the proposed MEG experiment at PSI making this class of
models testable in the near future.
\end{abstract} \maketitle
\vskip1.0in
\newpage

\section{Introduction}

Evidences for neutrino oscillations and hence for neutrino masses and
mixings seem to be getting more and more solid. The results
from solar and atmospheric neutrino experiments as well as those using
terrestrial neutrinos as in the K2K and KAMLAND experiments which are
responsible for this very important conclusion have
together with CHOOZ and PALO-VERDE experiments
clearly established the mixing pattern among different neutrino
generations. It seems that two of the three
neutrino mixings i.e. $\theta_{12}$ and $\theta_{23}$ are large and the
third angle $\theta_{13}$ is small\cite{review}. We will follow the
literature and call this the bi-large mixing pattern.

This good news on the mixing front is however not shared by the masses
i.e. our knowledge of the neutrino mass pattern is far from clear. At the
moment three obvious possibilities i.e. (i) normal mass hierarchy,
(ii) inverted mass hierarchy and (iii) degenerate mass pattern are equally
viable. The challenge for future experiments is therefore twofold: to pin
down the mass pattern more definitively and to improve measurements of the
three angles to more precise values so that different theoretical models
can be tested. For example it has been noted that if the present upper
limit on the angle $\theta_{13}$ drops by one order of magnitude, a lot of
very interesting models for neutrino masses will be ruled out. Of course
more models being ruled out makes the direction of physics beyond the
standard model that much more clear. It is therefore important to
extract the predictions of the various models as far as possible to narrow
the choices.

One of the very intriguing mass patterns for neutrinos is the inverted
pattern, where the two heavier neutrinos are split by the solar mass
difference $\Delta m^2_{\odot}$, whereas their absolute mass value is the
square root of
the atmospheric mass difference square i.e. $\Delta m^2_A$. In the
extreme limit where $\Delta m^2_{\odot}=0$, this mass matrix
admits an $L_e-L_\mu-L_\tau$ symmetry. If this pattern is confirmed
by experiments, the existence of the $L_e-L_\mu-L_\tau$ symmetry could
provide a clue to new local symmetries that operate at the higher scales
where neutrino masses arise.

In two recent papers\cite{kuchi}, it was shown that if there is a
leptonic $SU(2)_H$ horizontal symmetry\cite{other} acting
both on the left as well as the right handed leptons of the standard
model, then the theory must have two right handed neutrinos to cancel the
global $SU(2)_H$ Witten anomaly. This provides a rationale for the
existence of the right handed neutrino needed for implementing the seesaw
mechanism. This leads to a $3\times 2$ seesaw formula for neutrino masses. 
The resulting neutrino mass matrix naturally
leads to an inverted mass pattern for the light neutrinos after seesaw
mechanism\cite{seesaw} and in a certain areas of the parameter space gives
rise to an approximate $L_e-L_\mu-L_\tau$ symmetry\cite{emutau} of the
neutrino mass matrix. Thus if the inverted mass pattern for neutrinos is
confirmed by experiments, the $SU(2)_H$ models must be given very
serious consideration.

As discussed in ref.\cite{kuchi}, there are two possible patterns for the
charged lepton mass matrix in these models and since the PMNS mixing
matrix
gets contributions both from the charged lepton as well as the neutrino
sector, there will be two sets of predictions for the neutrino mixings. In
this paper, we focus on them separately and study their physical
implications, especially their predictions for lepton flavor violating
branching ratios $BR(\mu\rightarrow e+\gamma)$ and $BR(\tau\rightarrow
e,\mu+\gamma)$.

The key point is that below the horizontal symmetry scale, the theory is
the minimal supersymmetric standard model (MSSM) and
as is well known\cite{lfv1,lfv2,lfv3,bdm}, the presence of lepton mixings
leads to mixings between sleptons via the renormalization group evolution
even though at the GUT scale all slepton masses being equal forbid any
such mixings. These in turn lead to lepton
flavor violating decays such
as $BR(\mu\rightarrow e+\gamma)$ and $BR(\tau\rightarrow
e,\mu+\gamma)$ through one loop graphs involving the gauginos.

In order to arrive at our result, we first fix the model parameters such
that they lead to observed neutrino mixings.
 We then find that for both the models referred to above, in most of the
allowed parameter space for MSSM, the
 $BR(\mu\rightarrow e+\gamma)$ is around $10^{-12}$, which is within the
range of the next generation of experiments\cite{psi}. The constraints on
the supersymmetry parameter space come not only from the collider searches
for supersymmetry and the LEP bound on Higgs boson mass but also from the
Brookhaven g-2 results as well as the WMAP dark matter
constraints\cite{wmap}. The last constraint arises from the lightest
neutralino $\chi^0_1$ as the dark matter candidate and using the recent
WMAP results for the dark matter content of the universe i.e. $0.094<\Omega
h^2<0.129$ (2$\sigma$). In deriving these results, some assumptions are
needed for supersymmetry breaking parameters of the model. We use the
mSUGRA choice\cite{sugra}
 at the GUT scale to obtain the constraints i.e. universal mass for
superpartners, common gaugino mass and $A$-terms being proportional to
the Yukawa couplings. It is important that the GUT scale is higher than
the seesaw scale to get sizable flavor mixing effects.

This paper is organized as follows: in sec. 2, we briefly review the
salient features of the $SU(2)_H$ models and write down the $3\times 2$
seesaw
formulae for the neutrino masses and mixings as well as the charged lepton
mass matrices. We also present the exact neutrino mixing matrices in the
limit of zero electron masses. In sec. 3, we present the numerical
analysis of the neutrino mixings and masses and write down the choices of
the model parameters that are phenomenologically acceptable. In sec. 4, we
discuss the predictions for lepton flavor violating branching ratios in
the model for these choices of parameters. In sec. 5, we conclude with a
summary of the results.

 \section{Models for lepton mixing}

We consider extensions of the MSSM where the
gauge group extended to $SU(3)_c\times SU(2)_L\times U(1)_Y\times
SU(2)_H$ above the seesaw scale. Below the
seesaw scale the model is MSSM so that the usual results for gauge
coupling unification and electrweak symmetry breaking hold.

The basic motivation for considering these models is to find a symmetry
rationale for the existence of right handed neutrino othet than the fact
that it is needed to implement the
seesaw mechanism. An elegant extension of the standard model that brings
in the right handed neutrino automatically is to include the local B-L
symmetry in the theory i.e. theories based on the gauge group
$SU(2)_L\times U(1)_{I_{3R}}\times U(1)_{B-L}$ or $SU(2)_L\times
SU(2)_{R}\times U(1)_{B-L}$ or the GUT group SO(10). Since our
understanding of neutrino physics is not complete, it is important to
explore other possibilities.

A different possibility is to look for horizontal (or family) symmetries
that arise in the standard model in the limit of vanishing fermion masses.
To see the possibilities, let us work with the extension of the standard
model based on the gauge group $G_{STD}\times G_H$, where $G_H$ actos on
the different generations. Three interesting possibilities that provide
independent motivations for
the existence of right handed neutrinos
are: (i) $G_H=SU(3)_{q+\ell}$\cite{kribs}; (ii) $G_H=SU(2)_{q+\ell}$
\cite{kuchi}and
(iii) $G_H=SU(2)_{\ell}$\cite{kuchi} with the obvious choice of fermion
assignments. Here we will work with the case (iii) where
we assume that the quark superfields transform under the $SU(2)_{H}$ group
as singlets and leptons as doublets. In this case, it was noted in
\cite{kuchi} that while this model is free of local anomalies,
absence of global Witten anomalies\cite{witten}
implies that one must add two right handed neutrinos to the theory. Thus
the existence of the right handed neutrinos are required for the
consistency of the theory and not just to implement the seesaw mechanism.
This is the model we want to study in detail in this paper.

The assignment of the leptons and Higgs superfields
under the gauge group $ SU(3)_c\times SU(2)_L\times U(1)_Y \times SU(2)_H$
is given as follows:

\begin{center}
{\bf Table I}

\begin{tabular}{|c||c|}\hline
$\Psi \equiv (L_e, L_{\mu})$ & (1,2,-1,2)\\ \hline
$L_{\tau}$ & (1,2,-1,1) \\ \hline
$E^c \equiv (\mu^c, -e^c)$ & (1,1,-2, 2)\\ \hline
$\tau^c$ & (1,1,-2, 1)\\ \hline
$N^c\equiv (\nu^c_{\mu}, -\nu^c_{e})$ & (1,1,0,2)\\ \hline
$\nu^c_{\tau}$& (1,1,0,1)\\ \hline
$\chi_H\equiv \left(\begin{array}{cc} \chi_{1} & \chi_2 \\
\end{array}\right) $& (1, 1, 0, 2)\\ \hline
$\bar{\chi}_H\equiv \left(\begin{array}{cc} -\bar\chi_{2} & \bar\chi_1 \\
\end{array}\right)$ &
(1,1,0,2)\\\hline
$H_u$ & (1,2,1,1)\\ \hline
$H_d$ & (1,2,-1,1)\\ \hline
$\Delta_H$ & (1,1,0,3)\\\hline
\end{tabular}
\end{center}

\bigskip

\noindent{\bf Table caption: } We display the quantum number of the matter
and Higgs superfields of our model.

Here $L_{e,\mu,\tau}$ denote the left handed lepton doublet superfields.
We arrange the Higgs potential in such a way that the $SU(2)_H$ symmetry
is broken by $<\chi_1>=<\bar{\chi}_2>=
v_{H1}; <\chi_2>=<\bar{\chi}_1>=v_{H2}$ and $<\Delta_{H,3}>=
v'_H$, where $v_{H,i},v'_H \gg v_{wk}$. The vevs for $\bar{\chi}$ are
chosen
so as
to cancel the D-terms and leave supersymmetry unbroken below the scale
$v_H$. Note that we have used the $SU(2)_H$
symmetry to align the $\Delta_H$ vev along the $I_{H,3}$ direction.
At the weak scale,
all the neutral components of the fields $H_u$ and $H_d$ acquire
nonzero vev's and break the standard model symmetry down to $SU(3)_c\times
U(1)_{em}$. We denote these vev's as follows: $<H^0_{u}>
=\kappa_{0}$ and $<H^0_d> = \kappa_0 cot\beta$; Clearly $\kappa_0$
has values in the 100 GeV range depending on the value of $tan\beta$.

Note that $<\Delta_H>\neq 0$ breaks
the $SU(2)_H$ group down to the $U(1)_{L_e-L_{\mu}}$ group which is
further broken down by the $\chi_H$ vevs. Since the renormalizable
Yukawa interactions do not involve the $\chi_H$ field, this symmetry
($L_e-L_{\mu}$) is also reflected in the right handed neutrino mass
matrix. This turns out to play a crucial role in leading to a bimaximal
mixing pattern.

To discuss the detailed leptonic mixings, we now write down the
Yukawa superpotential of the theory. The two models we discuss arise from
the two ways the horizontal doublets couple to fermion superfields. In one
case, we will allow only the $\chi_H$ field to couple to all fermions (to
be called model I) and in the second
case, we will allow the $\chi_H$ to couple to the right handed neutrino
sector only and $\bar{\chi}_H$
to couple to the charged leptons only. This will be called model II.

\subsection{Model I}

The Yukawa superpotential for this model is given by:
 \begin{eqnarray}
 W_Y&=&h_0 (L_eH_u\nu^c_e+L_\mu H_u\nu^c_\mu)
+h_1 L_\tau(\nu^c_\mu \chi_2 + \nu^c_e\chi_1)H_u/M
-if N^{c T}\tau_2{\bf \tau \cdot \Delta_H}N^c \\ \nonumber
&&\frac{h'_1}{M}(L_e\chi_2-L_\mu \chi_1)H_d\tau^c+
\frac{h'_4}{M} L_\tau H_d(\mu^c\chi_2+e^c\chi_1)+
{h'_3}L_{\tau}H_d \tau^c
+h'_2 (L_ee^c+L_\mu \mu^c)H_d
\end{eqnarray}
This can be made natural by a choice of the discrete $Z_2$ symmetry
under which $\bar{\chi_H}$ is odd and all other fields even, so
that it does not couple to matter fields in the superpotential.
$<\Delta^0_H>= v'_H$ directly leads to the $L_e-L_{\mu}$ invariant
  $\nu_{eR}-\nu_{\mu R}$ mass matrix at the seesaw scale, as already
noted. The $\chi_H$ vev contributes to this mass matrix only through
nonrenormalizable operators and we assume those contributions to be
negligible. We also do not include any term where $\Delta_H$ couples to
light fields. Since the theory is supersymmetric, any term omitted from
the superpotential will not be induced by loops due to the
non-renormalization theorem.
Similarly there will also be some small contributions from the
$\nu_{\tau R}$ sector if we did not decouple it completely. We ignore
these contributions in our analysis. Further, we define
$\kappa_{1,2}= \frac{<\chi_{1,2}>\kappa_0}{M}\simeq 1-100$ GeV.

 To study neutrino mixings in this model, let us write
down the $5\times 5$ mass matrix for neutrinos:
\begin{eqnarray}
M_{\nu_L,\nu_R}~=~\left(\begin{array}{ccccc} 0 & 0 & 0 & h_0\kappa_0 & 0\\
0 & 0 & 0 & 0 & h_0\kappa_0\\ 0 & 0 & 0 & h_1\kappa_1 & h_1 \kappa_2 \\
h_0\kappa_0 & 0 & h_1\kappa_1 & 0 & fv'_H  \\ 0 & h_0\kappa_0 & h_1\kappa_2
& fv'_H & 0 \end{array} \right)
\end{eqnarray}
This is a $3\times 2$ seesaw matrix\cite{kuchi,fram} and
after seesaw diagonalization, it leads to a $3\times 3$ light neutrino
mass matrix as follows:
\begin{eqnarray}
{\cal M}_{\nu}~=~-M_D M^{-1}_R M^T_D
\end{eqnarray}
where $M_D~=~\left(\begin{array}{cc} h_0\kappa_0 & 0 \\ 0 & h_0\kappa_0\\
h_1\kappa_1 & h_1\kappa_2
\end{array}\right)$; $M^{-1}_R~=~\frac{1}{fv'_H}\left(\begin{array}{cc} 0
&
1\\1 & 0 \end{array}\right)$. The resulting light Majorana neutrino mass
matrix ${\cal M}_{\nu}$ is given by:
\begin{eqnarray}
{\cal M}_{\nu}~=~-\frac{1}{fv'_H}\left(\begin{array}{ccc} 0 &
(h_0\kappa_0)^2 & h_0h_1\kappa_0\kappa_2\\ (h_0\kappa_0)^2 & 0 &
h_0h_1\kappa_0\kappa_1 \\ h_0h_1\kappa_0\kappa_2 & h_0h_1 \kappa_0\kappa_1
& 2h^2_1\kappa_1\kappa_2 \end{array}\right)
\label{mnu}
\end{eqnarray}
To get the physical neutrino mixings, we also need the charged lepton mass
matrix defined by $\bar{\psi}_L {\cal M}_\ell \psi_R$. This is where the
difference between the two models appear.

In model I, the charged
lepton mass matrix is given by
\begin{eqnarray}
{\cal M}_{\ell}~=~cot\beta \left(\begin{array}{ccc} h'_2\kappa_0
& 0
&-h'_1\kappa_2
\\ 0 & h'_2\kappa_0 & h'_1\kappa_1 \\ h'_4\kappa_1
& h'_4\kappa_2 &
h'_3\kappa_0 \end{array}\right).
\end{eqnarray}
 In order to study physical neutrino
mixings, we must diagonalize the
${\cal M}_{\nu}$ and $M_\ell$ matrices. In ref.\cite{kuchi}, we
diagonalized the neutrino mass matrix analytically in the limit
$m_e=0$. Let us discuss it for
completeness. Since $h'_2\kappa_0$ is proportional to the electron mass,
we will ignore temporarily. Then,
defining the matrices that diagonalize the charged lepton mass matrix as
$D_\ell = U^{(L)}_{\ell} M_{\ell} U^{(R)\dagger}_{\ell}$, we get
\begin{eqnarray}
U^{(L)}_{\ell}=\left(\begin{array}{ccc}
s_1 & c_1 & 0\\c_\beta c_1 & -c_{\beta}s_1 & s_\beta \\
-s_\beta c_1 & s_\beta s_1 & c_\beta \end{array}\right)
\end{eqnarray}
where two angles $\theta_{1,2}$ are given by:
\begin{eqnarray}
sin \theta_1 \equiv s_1 = \frac{\kappa_1}{\sqrt{\kappa_1^2+\kappa_2^2}}\\
\nonumber
sin \theta_2 \equiv s_2 = \frac{h_0\kappa_0}{\sqrt{h^2_0\kappa^2_0+
h^2_1(\kappa^2_1 +\kappa^2_2)}} \\ \nonumber
\end{eqnarray}
and $tan 2\beta \simeq 2\sqrt{\frac{h'_1}{h'_4}} \sqrt
{m_{\mu}/m_{\tau}}$ and we have ignored terms of order $m_e/m_\mu$.
Similarly the matrix that diagonalizes ${\cal M}_{\nu}$ is given by
\begin{eqnarray}
U_{\nu}~=~\left(\begin{array}{ccc} c_1c'-s_1s_2s' & c_1s'+s_1s_2c' &
c_2s_1 \\
-s_1c'-c_1s_2s' & -s_1s'+c_1s_2c' & c_2c_1 \\ -c_2s' & c_2c' & -s_2
\end{array}\right)
\end{eqnarray}
where $s'= sin\theta'$ with $\theta'$ given by
$tan2\theta'=\frac{2s_2(c^2_1-s^2_1)}{(1+s^2_2)2s_1c_1}$.
The PMNS matrix that is measured in neutrino oscillation is given by
$U^{(L)}_\ell U_{\nu}$ and has the form, in the limit of $m_e=0$:
\begin{eqnarray}
{\bf \Large U}_{PMNS}=\left(\begin{array}{ccc} s_2s' & s_2c' & c_2 \\
-(c'c_\beta+c_2s's_\beta) & -(c_\beta s'-s_\beta c_2c') & -s_2s_\beta \\
(c's_\beta - c_2 s' c_\beta) & ( c_2c'c_\beta+s's_\beta) & s_2 c_\beta
 \end{array}\right)
\end{eqnarray}
Using this, it was concluded in \cite{kuchi} that
there is an approximate relation between the neutrino parameters given by:
\begin{eqnarray}
U^2_{e3} cos 2\theta_{\odot} \simeq \frac{\Delta m^2_{\odot}}{ 2 \Delta
m^2_A}
\end{eqnarray}
It is clear from the above equation that for a given value of $\Delta
m^2_{\odot}/\Delta m^2_A$, the smallest value of $U_{e3}$
will happen for the largest value of $sin^22\theta_{\odot}$.
In this paper first we numerically solve for the neutrino mixings and then
proceed to calculate the lepton flavor violation effects for the allowed
choices of the parameters of the model.

\subsection{Model II}
In this case, we impose a discrete symmetry under which the fields
$(e^c,\mu^c,\tau^c)$ and $\bar{\chi}_H$ are odd and all othee fields are
even. The Yukawa superpotential invariant under this is:
 \begin{eqnarray}
 W_Y~=~h_0 (L_eH_u\nu^c_e+L_\mu H_u\nu^c_\mu)
+h_1 L_\tau(\nu^c_\mu \chi_2 + \nu^c_e\chi_1)H_u/M
-if N^{c T}\tau_2{\bf \tau \cdot \Delta_H}N^c +\\ \nonumber
\frac{h'_1}{M}(L_e\bar\chi_1+L_\mu \bar\chi_2)H_d\tau^c+
\frac{h'_4}{M} L_\tau H_d(\mu^c\bar\chi_1-e^c\bar\chi_2)+
{h'_3}L_{\tau}H_d \tau^c
+h'_2 (L_ee^c+L_\mu \mu^c)H_d
\end{eqnarray}
Note that the vev's $<\bar{\chi}_{1,2}>= \kappa_{2,1}$ for the D-terms to
cancel at the high scale. Substituting these vevs and the vev's of
$H_{u,d}$, we get exactly the same neutrino mass matrix as in model I
i.e. Eq.\ref{mnu} but the charged lepton mass matrix becomes:
\begin{eqnarray}
{\cal M}_{\ell}~=~cot\beta \left(\begin{array}{ccc} h'_2\kappa_0
& 0
&h'_1\kappa_1
\\ 0 & h'_2\kappa_0 & h'_1\kappa_2 \\ -h'_4\kappa_2
& h'_4\kappa_1 &
h'_3\kappa_0 \end{array}\right).
\end{eqnarray}
The matrices that diagonalize this matrix are different from the first
case (model I) and the $U^{(L)}_\ell$ in this case is given by:
\begin{eqnarray}
U^{(L)}_{\ell}=\left(\begin{array}{ccc}
c_1 & -s_1 & 0\\c_\beta s_1 & c_{\beta}c_1 & s_\beta \\
-s_\beta s_1 & -s_\beta c_1 & c_\beta \end{array}\right)
\end{eqnarray}
The ${\bf U}_{PMNS}$ in this case is given by
\begin{eqnarray}
{\bf \Large U}_{PMNS}=\left(\begin{array}{ccc} c' & s' & 0 \\
-s's_2c_\beta-c_2s's_\beta & s_2c'c_\beta+c_2c's_\beta &
c_2c_\beta-s_2s_\beta\\
(-s's_2s_\beta - c_2 s' c_\beta) & ( c_2c'c_\beta-s_2c's_\beta) & -s_2
c_\beta-c_2s_\beta
 \end{array}\right)
\end{eqnarray}
Note that we have neglected corrections of order $\sqrt{m_e/m_\mu}$ and
once they are turned on, we indeed have a nonzero $U_{e3}$ and small.
We will discuss the detailed numerical predictions in the next section.

\section{Neutrino mass fit}
\subsection{MODEL I}
The expressions for the Dirac neutrino matrix and the charged lepton
matrix are given in the previous section. We
will use those expressions in order to derive the light neutrino masses
and the mixing angles numerically. That will fix the model parameters
that will go into the renormalization group evolution of the slepton
masses. We will
generate the fit for different values of $\tan\beta$ and  model parameters.
For example, at $\tan\beta=30$, the charged lepton mass matrix is
 given by (at the scale where the horizontal symmetry gets
broken) :
\begin{eqnarray} {\cal M}_{\ell}&=&\left(\matrix{
  7.5\times 10^{-3} & 0 & -1.135\cr
  0 & 7.5\times 10^{-3} & 0.247\cr
 2.14\times 10^{-2} & 0.0984& 0.85  }\right).
\end{eqnarray}
 This matrix needs $h_1'=1.135\times 10^{-1}$, $h_2'=1.29\times 10^{-3}$,
$h_3'=1.466\times 10^{-1}$ and  $h_4'=9.8\times 10^{-3}$ with
$\kappa_2=10$ and ${\kappa_2\over \kappa_1}=4.588$.
We get  the correct values of charged lepton masses at the weak scale
from the above matrix. We use the MSSM RGEs between the horizontal scale
and the
weak scale. Using the same $\kappa$s, $h_0=9.2\times 10^{-2}$ and
$h_1=-2.92\times 10^{-1}$, we find the light neutrino mass matrix (in eV):
\begin{eqnarray}  {\cal M}_{\nu}&=&\left(\matrix{
  0 & 5.15\times 10^{-2} & -9.3\times 10^{-3}\cr
  5.15\times 10^{-2} & 0 & -2.04\times 10^{-3}\cr
 -9.3\times 10^{-3}& 2.04\times 10^{-3}& 7.4\times 10^{-4}  }\right).
\end{eqnarray}
The PMNS matrix and neutrino eigenvalues at the weak scale are :
\begin{eqnarray} {\bf \Large U}_{PMNS}&=&\left(\matrix{
 0.806 & 0.58 & -0.109 \cr
  -0.27& 0.527 & 0.805\cr
  0.526& -0.619& 0.582}\right).
\end{eqnarray}
\begin{equation}
(m_1,~m_2,~m_3) = (5.276\times
10^{-2},\, 5.202\times 10^{-2},\,  0)~ {\rm eV}~.
\end{equation}
From the above fit, we find $\sin^22\theta_{\odot}=0.87$, $\Delta
m^2_{\odot}=7.7\times 10^{-5}$ eV$^2$ and
$\sin^22\theta_{A}=0.88$, $\Delta m^2_A=2.7\times 10^{-3}$ eV$^2$ which
are well
allowed by the experimental results.

The above fit is obtained for a particular value of $\kappa_2\over
\kappa_1$. We can change that value and the  Yukawa
couplings $h$s and $h'$s and still can obtain very good fits. In fig.1 we
show the model points which are within the experimental
limits of $\sin^22\theta_{\odot}$ and $\sin^22\theta_{A}$. We find many
model points which satisfy the
allowed ranges. In Fig.2, we show the $|U_{e3}|$ as a function of
$\sin^22\theta_{\odot}$ for the model points
which satisfy the experimental limit on $\sin^22\theta_{A}$. The range of
$U_{e3}$
is 0.07-0.13 in this model and is allowed by the current experimental
upper bound 0.22\cite{fl}. In fact, it is expected that the planned long base
line experiments including MINOS have the ability to probe this range in
the next few years.

The scale of the horizontal symmetry breaking is assumed to be $3\times
10^{13}$ and the magnitude of the Majorana coupling
is $f\sim 1.6\times 10^{-1}$. The fit does not change if we move the
scale up or down.

The fit for $\tan\beta=$10 and 40 look similar. The charged lepton mass
matrices for these two $\tan\beta$ are
given below.\\
For $\tan\beta=40$ we have:\begin{eqnarray} {\cal M}_{\ell}&=&\left(\matrix{
  8.2\times 10^{-3} & 0 & -1.30\cr
  0 & 8.2\times 10^{-3} & 2.83\times 10^{-1}\cr
 2.33\times 10^{-2}& 1.07\times 10^{-1}& 9.5\times 10^{-1}  }\right).
\end{eqnarray} We keep the light neutrino matrices unchanged and we find
$\sin^22\theta_{\odot}=0.88$, $\Delta m^2_{\odot}=7.7\times 10^{-5}$ eV$^2$ and
$\sin^22\theta_{A}=0.87$, $\Delta m^2_A=2.7\times 10^{-3}$ eV$^2$.\\
For  $\tan\beta=10$ we have:\begin{eqnarray} {\cal M}_{\ell}&=&\left(\matrix{
  6.8\times 10^{-3} & 0 & -9.9\times 10^{-1}\cr
  0 & 6.8\times 10^{-3} & 2.15\times 10^{-1}\cr
 2.088\times 10^{-2}& 9.58\times 10^{-2}& 8.0\times 10^{-1}  }\right).
\end{eqnarray} We keep the light neutrino matrices unchanged and we find
$\sin^22\theta_{\odot}=0.88$, $\Delta m^2_{\odot}=7.7\times 10^{-5}$ eV$^2$ and
$\sin^22\theta_{A}=0.92$, $\Delta m^2_A=2.7\times 10^{-3}$ eV$^2$.
\begin{figure}\vspace{0cm}
    \begin{center}
    \leavevmode
    \epsfysize=8.0cm
    %\epsffile[75 160 575 630]{/home/duttabh/lfv/neutrinofit2.eps}
    \epsffile[75 160 575 630]{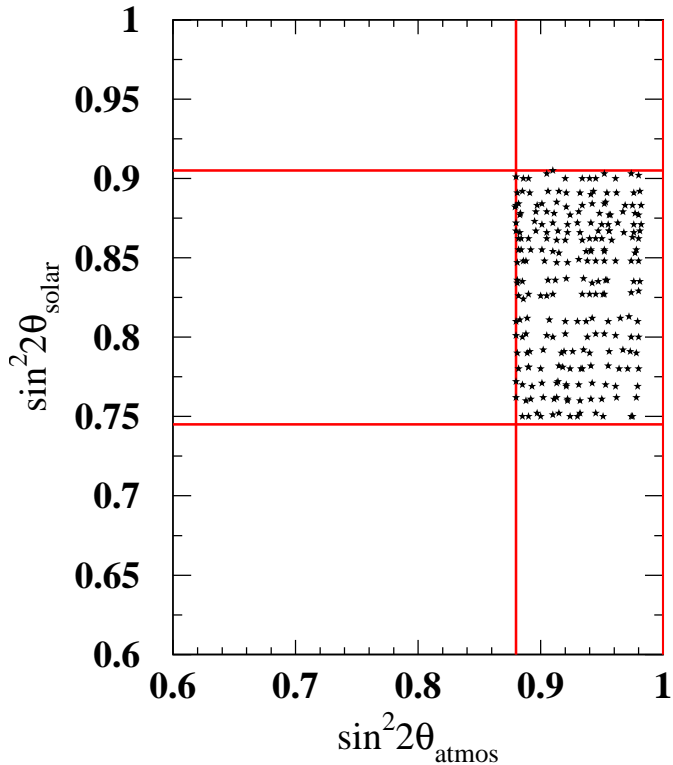}
    \vspace{2.0cm}
     \caption{\label{fig:fig1} $sin^22\theta_{\odot}$ vs
$sin^22\theta_{\rm atmos}$. The stars show the model
    points.  The horizontal and vertical
    solid lines show the current experimental limits.}
\end{center}\end{figure}

\begin{figure}\vspace{0cm}
    \begin{center}
    \leavevmode
    \epsfysize=8.0cm
    %\epsffile[75 160 575 630]{/home/duttabh/lfv/neutrinofit3.eps}
    \epsffile[75 160 575 630]{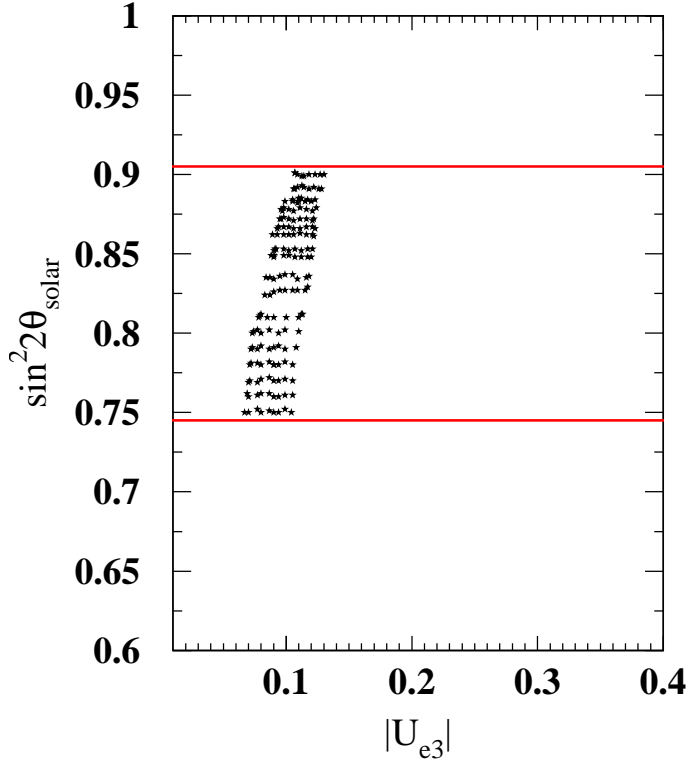}
    \vspace{2.0cm}
     \caption{\label{fig:fig2} $sin^22\theta_{\odot}$ vs $|U_{e3}|$.  The
stars show the model
    points.  The horizontal
    solid lines show the  current experimental limits.}
\end{center}\end{figure}
\subsection{MODEL II}
In this model, the Dirac neutrino matrix is similar to what we have in
the previous model. The charged lepton
matrix is the transpose of that in the other case. We can again obtain
the neutrino fit for different values of
$\tan\beta$.
For example, at $\tan\beta=30$, The charged lepton mass matrix  is
given by (at the scale where the horizontal symmetry breaks
down):\begin{eqnarray} {\cal M}_{\ell}&=&\left(\matrix{
  8.28\times 10^{-2} & 0 & 5.0\times 10^{-5}\cr
  0 & 8.28\times 10^{-2} & 4.0\times 10^{-3}\cr
 -1.44& 0.018& 6.9\times 10^{-3}  }\right).
\end{eqnarray}
This matrix requires $h_2'$=0.0143, $h_3'=1.2\times 10^{-3}$,
$h_1'=1.00\times 10^{-4}$,  $h_4'=3.6\times 10^{-2}$ for
$\kappa_2=40$ and ${\kappa_2\over \kappa_1}=0.5$.
The charged lepton mass matrix produces the correct values of charged
lepton masses at the weak scale after using the RGEs.
Using the same $\kappa$'s, $h_0=7.67\times 10^{-2}$ and $h_1=3.07\times
10^{-1}$ we find the light neutrino mass matrix:
\begin{eqnarray} {\cal M}_{\nu}&=&\left(\matrix{
  0 & 3.55\times 10^{-2} & 3.27\times 10^{-2}\cr
  3.55\times 10^{-2} & 0 & 4.09\times 10^{-4}\cr
 3.27\times 10^{-2}& 4.09\times 10^{-4}& 7.5\times 10^{-4}  }\right).
\end{eqnarray}
The PMNS matrix and neutrino eigenvalues are :
\begin{eqnarray} {\Large\bf U}_{PMNS}&=&\left(\matrix{
 -0.724 & 0.68 & -0.042 \cr
  0.527& 0.513 & -0.677\cr
  -0.44& -0.512& -0.734}\right).
\end{eqnarray}
\begin{equation}
(m_1,~m_2,~m_3) = (4.87\times
10^{-2},\, 4.79\times 10^{-2},\,  0)~ {\rm eV}~.
\end{equation}
The solar angle prediction i.e. $sin^22\theta_{\odot}=0.95$ for this
model is on the higher side and
therefore as the experimental error on $\theta_{12}$ narrows, this
model will be tested. The atmospheric mixing angle
$sin^22\theta_A$ is 0.98 and is well allowed by the experimental
results. The prediction for $U_{e3}=-0.042$ and it will also be probed in
the near future.

In the following, the scale of horizontal symmetry breaking our
calculations is assumed to
be $3\times 10^{13}$ and the magnitude of the Majorana coupling
is $f\sim1.6\times 10^{-1}$.

\section{Lepton Flavor Violation}

It is well known that in the standard model where neutrino masses vanish,
there is no lepton flavor violation. Once the neutrinos have masses as
well as mixings, there are small amounts of lepton flavor violation
leading to processes such as $\ell_j\rightarrow \ell_i + \gamma$ with
$i\neq j$. However, if only a Dirac mass( by adding a right handed
neutrino) or a Majorana mass (by adding a Higgs triplet) of the neutrino
are added to the standard model, then lepton flavor violating
processes which always arise at one loop or higher loop level in gauge
theories have branching ratios proportional to
$\left(\frac{m^2_\nu}{m^2_W}\right)^2$ and are therefore very tiny. In
models with supersymmetry however, the situation can change
drastically. In
the presence of neutrino masses, the superpartners of leptons could in
general mix and the LFV branching ratio in that case becomes of
order$\left(\frac{m^2_{\tilde{\ell}}}{m^2_W}\right)^2$ where
$m_{\tilde{\ell}}$ is a typical mass of the slepton and since the
slepton masses are expected to be of order of the weak scale, the LFV
branching ratios in general can be quite large. The actual magnitude
however depends on the specific model for neutrino masses as well as on
the pattern of supersymmetry breaking. If we assume the neutrino masses
to arise from the seesaw mechanism, then depending on how the seesaw
mechanism is implemented in a theory, the predictions for LFV can be
different.

Similarly, the nature of supersymmetry breaking is also important since
in the supersymmetric limit the radiative flavor changing decays vanish.
To illustrate this point, let us consider a simple seesaw model where
the susy breaking slepton masses are universal at the seesaw scale and the
SUSY breaking $A$ terms are
proportional to Yukawa couplings as in the minimal SUGRA models. We can
go to a basis where the charged lepton Yukawa couplings are diagonal. This
simply rotates the neutrino mass matrix and keeps all other terms in the
Lagrangian flavor  In such a theory, just below the
seesaw scale, flavor mixing is present only in the neutrino masses from
the seesaw
formula. As we extrapolate the slepton masses down to the seesaw scale,
they remain diagonal since the charged lepton Yukawa coupling is diagonal.
The only source of lepton flavor violation is then in the neutrino
masses. Therefore, in this case
$BR(\mu\rightarrow e+\gamma) \propto
\left(\frac{m^2_\nu}{m^2_W}\right)^2$ and is therefore negligible.
To obtain a significant LFV effect therefore, one must assume that the
universality of slepton masses and the proportionality of the $A$ terms
must hold much above the seesaw scale. Fortunately, this is what is
generally assumed in minimal SUGRA models anyway.
Thus no observable LFV effect would simply be a manifestation of the fact
that it is the seesaw scale where universality of slepton masses holds
e.g. in models like gauge mediated supersymmetry breaking.
In this paper, we will assume that
the universality of scalar masses holds at a scale higher than the
seesaw scale.

Turning to our case, since we only have an SU(2) horizontal
symmetry, we can dispense with any assumption regarding universality and
allow for the most general SUSY breaking pattern consistent
with the $SU(2)_H$ symmetry. For definiteness, we will assume
universality of scalar masses at
a higher scale (e.g. the GUT scale) and use RGE effects of the lepton
Yukawa couplings to get the slapton masses and the $A$ terms at the $v_H$
scale. This is a non-universal profile at the seesaw scale and
leads to $m^2_i$ as a function of the Yukawa couplings. If we then rotate
i.e. $m^2_{\tilde{e}}=m^2_{\tilde{\mu}}=m^2\simeq
m^2_0+\frac{(6+2a^2)(h^2_0+h^{'2}_2)m^2_0}{16\pi^2}\ell n\frac{M_U}{v_H}$
and
$m^2_{\tilde{\tau}}=m^2_0
+\frac{(6+2a^2)(h^{'2}_3)m^2_0}{16\pi^2}\ell n\frac{M_U}{v_H}$. If
we now rotate
the charged lepton Yukawas just below the horizontal symmetry scale to go
to a diagonal basis for charged lepton Yukawa couplings, we will generate
slepton flavor mixings such as
$m^2_{LL,13}$ and $m^2_{LL,23}$. A combination of these two terms at the
weak scale will give rise to a nonvanishing branching ratio for
$\mu\rightarrow e +\gamma$ and $\tau\rightarrow e,\mu+\gamma$.
If we focus the slepton masses alone, we roughly get
\begin{eqnarray}
B(\mu\rightarrow e+\gamma)\propto \frac{\alpha_{em}\alpha^2_1\delta^2
(U_{e3}U_{\mu 3})^2}{G^2_Fm^4_0}
\end{eqnarray}
One can also in a similar manner estimate the contribution of the $A$
terms. In fact it turns out that it is the
$A$-term slepton mixings that make the dominant contribution to the
LFV branching ratios in our model.

To proceed with the calculations for LFV processes in our case, we will
use  mSUGRA models which has five parameters at the GUT scale i.e. $m_0$,
$m_{1/2}$, $A_0$, $\tan\beta$ and
 $\mu$\cite{sugra}. We will assume that at the GUT scale all scalar
masses have the common value $m_0$ as noted in the previous
section; gauginos have the common mass
$m_{1/2}$; the trilinear SUSY breaking terms are proportional to the
Yukawa couplings and the rest of the theory is defined by the
 the superpotentials in section II. First note that the renormalization
group evolution (RGE) is caused only by dimension four terms in the
Lagrangian.
 The charged lepton Yukawa matrix responsible for the RGE from the GUT
scale to the weak scale is diagonal since
the only dimension four Yukawa couplings are $h'_{2,3}$.
Similarly, the dimension 4  Dirac neutrino couplings are also flavor
diagonal with the Yukawa coupling
$h_0$. Thus as we run down from the GUT scale to the horizontal symmetry
breaking scale $v_H$, the slepton masses remain horizontal; The $A$ terms
get modified.

The RGE from the $v_H$ scale down to the weak scale is the MSSM RGE.
We generate the off diagonal elements of the charged lepton Yukawa matrix
at the horizontal scale from the nonrenormalizable operators and Dirac
neutrino matrix becomes a $3\times2$ matrix.  The off diagonal elements
of the charged lepton mass matrix induce the flavor violating
terms in the $m_{LL}^2$,
 $m_{LR}^2$ and $m_{RR}^2$ mass matrices via the RGEs from the horizontal
scale down to the weak scale.
 For example, $E_{A_{12}}$ term (which contributes to $m_{LR}^2$) gets
generated via terms $ E_A U_l^{\dag} U_l$, $U_lU_l^{\dag} E_A$ pieces in
the RGE
 (${dE_A\over {dt}}$$\propto {1\over {16 \pi^2}}(4 E_A U_l^{\dag} U_l+
 5 U_lU_l^{\dag} E_A)$, where $U_l$ is the $3\times 3$ matrix for the Dirac
Yukawa coupling). We have $U_{13,31}$ and $U_{32,23}$ terms in both
models. These terms  generate large BR($\mu\rightarrow e+\gamma$).

 The calculation of BR($\mu\rightarrow e+\gamma$)involves both neutralino
and chargino diagram contribution. In some parameter
 space one type dominates and the other type dominates in some other.
 Before we discuss the BRs. of the lepton flavor violating decay modes,
we need to discuss the supersymmetric
 parameter space.

\subsection{Supersymmetry parameter space}
 The mSUGRA parameter space is constrained  by the experimental
lower limit on $m_h$ and measurements of $b\rightarrow s\gamma$ and recent
results on dark matter relic density\cite{wmap}. For low $\tan\beta$, the
parameter space has lower bound on  $m_{1/2}$ from the light Higgs mass
bound of $m_h\geq 114$ GeV. For larger $\tan\beta$ the lower bound on
$m_{1/2}$ is produced by the CLEO constraint on the BR($b\rightarrow
s\gamma$). The recent WMAP results have led to stronger constraints on dark
matter density which in turn reduces the allowed parameter space of MSSM
mostly to the co-annihilation region for $m_0,\,m_{1/2}\leq 1000$ GeV. As
is well known, co-annihilation requires some other superpartner to have
mass close to the lightest neutralino. In the mSUGRA model co-annihilation
happens between the lightest slepton and the lightest
neutralino\cite{coan}.
  One can see this analytically for low and intermediate
$\tan\beta$ for mSUGRA where the RGE can be solved
analytically~\cite{b35}. At the
electroweak scale one has for the average slepton masses
\begin{equation}
    \tilde m_{e_R}^2 = m_0^2  + (6/5) f_1 m_{1/2}^2 -\sin^2\theta_W M_W^2
    \cos(2\beta)     \label{eq14}
\end{equation}
\begin{equation}
   m_{\tilde{\chi}_{1}^{0}}= (\alpha_1/\alpha_G) m_{1/2}    \label{eq15}
\end{equation}
where $f_i = [1-(1+\beta_i t)^{-2}]/\beta_i$, $t= \ln(M_G/M_Z)^2$ and
$\beta_1$ is the $U(1)$ $\beta$
function. Numerically this gives for e.g. $\tan\beta=$ 5
\begin{eqnarray}
          \tilde m_{\tilde e_R}^2 &=& m_0^2\, + \,0.15 m_{1/2}^2\,  + (37\,{\rm
      GeV})^2 \nonumber \\
             m_{\tilde{\chi}_{1}^{0}}^2 &=& 0.16 m_{1/2}^2      \label{eq16}
\end{eqnarray}
Thus for $m_0=$ 0, the $\tilde e_R$ becomes degenerate with the
$\tilde{\chi}_{1}^{0}$ at $m_{1/2} = 370$ GeV,
i.e. coannihilation effects begin at $m_{1/2}\stackrel{\sim}{=}(350-400)$
GeV. For larger
$m_{1/2}$, the near degeneracy is maintained by increasing $m_0$, and
there is a
corridor in the $m_0-m_{1/2}$ plane  allowing for an adequate relic density.
For larger $\tan\beta$, the stau is the lightest slepton and its mass is
close to the neutralino mass for a large region of parameter space.

The previous relic density bound $0.07<\Omega h^2<0.21$ reduced the
parameter space mostly into coannihilation bands. The most recent WMAP result
$0.094<\Omega h^2<0.129$ has narrowed the co-annihilation regions
further\cite{wmap2}.

Further constraints on the parameter space arise from the present muon g-2
results\cite{ads}, if we take the deviation from the standard model value
as an upper limit on the supersymmetric contribution to g-2. For instance,
for $\tan\beta$=10, we get  $m_{1/2}< 430$ GeV.
Since the Higgs
mass implements a lower bound $m_{1/2}\geq $300 GeV, we do not have much
parameter space left for this value of $\tan\beta$.
However, for $\tan\beta$=30 and 40, the  upper bounds are $m_{1/2}< 720$
and 790 GeV respectively from the g-2 of muon data and thus the allowed parameter space
becomes bigger. We will
therefore restrict ourselves to larger values of $\tan\beta$ for the
discussions of flavor violation BRs.
\begin{figure}\vspace{0cm}
    \begin{center}
    \leavevmode
    \epsfysize=8.0cm
    \epsffile[75 160 575 630]{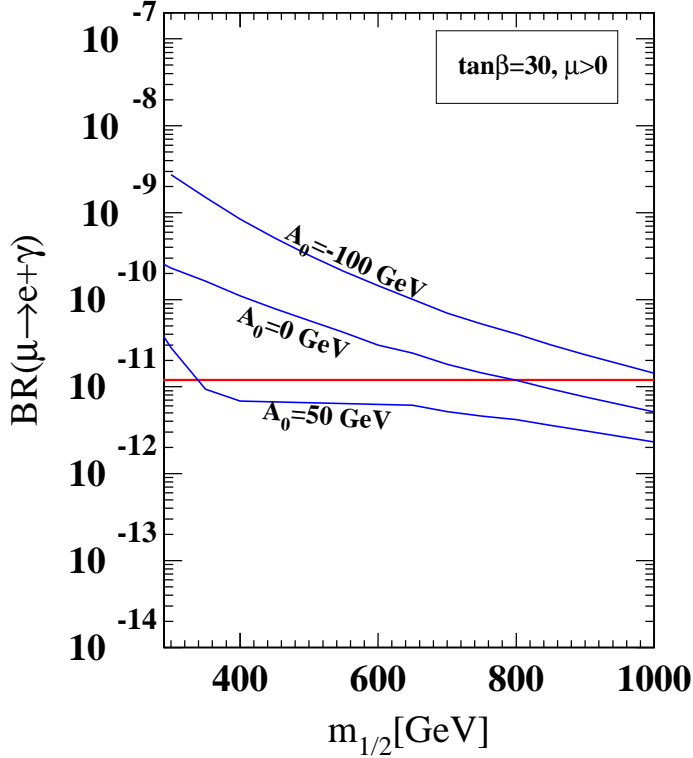}
    %\epsffile[75 160 575 630]{/home/duttabh/lfv/muegmh30.eps}
    \vspace{2.0cm}
     \caption{\label{fig:fig3} BR[$\mu\rightarrow e+\gamma$] as a function
of $m_{1/2}$ for different values of  $A_0$ for model I.}
\end{center}\end{figure}

\begin{figure}\vspace{0cm}
    \begin{center}
    \leavevmode
    \epsfysize=8.0cm
    \epsffile[75 160 575 630]{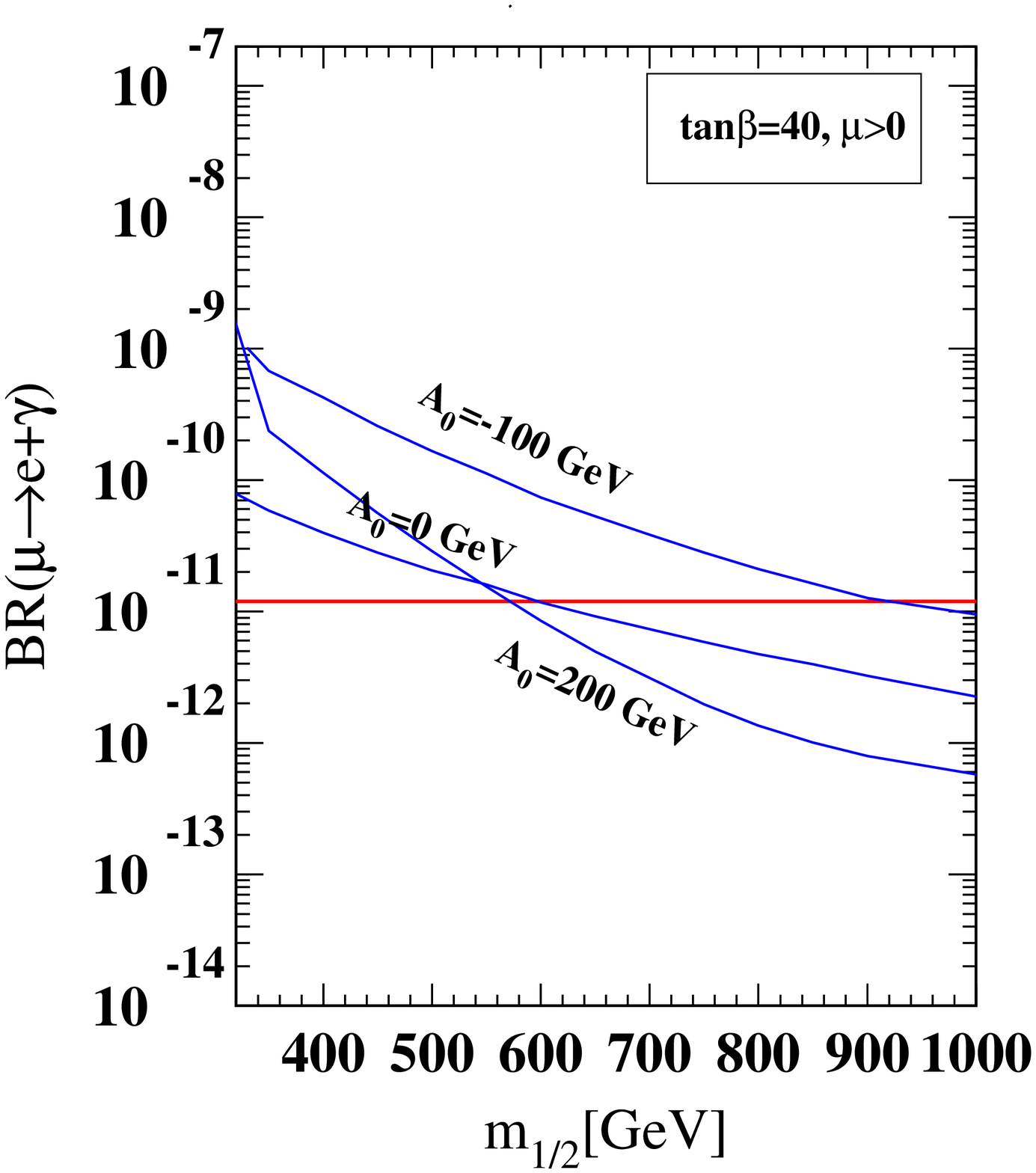}
    %\epsffile[75 160 575 630]{/home/duttabh/lfv/muegmh40.eps}
    \vspace{2.0cm}
     \caption{\label{fig:fig4} BR[$\mu\rightarrow e+\gamma$] as afunction
of $m_{1/2}$ for different values of  $A_0$ for model I.}
\end{center}\end{figure}
  In Figs.3 and 4, we show the BR($\mu\rightarrow e+\gamma$) for
$\tan\beta$=30 and 40 for different values of $A_0$
in model I.
 We tune $m_0$ such so that $\Omega h^2$ constraint is satisfied. This
 mostly happens where the light stau mass is very close to the lightest
neutralino mass.
 For example, for $m_{1/2}=400$ GeV, $m_0\sim$ 208-215 GeV satisfies the
$\Omega h^2$ constraint for $\tan\beta=40$ and $A_0=0$.
 The BR remains almost unchanged over the small range of $m_0$. The parameter
 space shown in the figures are allowed by the BR$[b\rightarrow s\gamma]$
 constraint.
%The left
%axis of the graphs is chosen to satisfy the
% BR[$b\rightarrow s\gamma$]  limit $2\times 10^{-4}$ for the given
% $\tan\beta$ for $A_0$=0. Positive $A_0$, e.g. $A_0=200$ GeV, pushes back
%the axis by 20 GeV (since the
% gain is small we are keeping the left axis fixed at the BR[$b\rightarrow
 %s\gamma$]  limit  for $A_0$=0) and the negative $A_0$ e.g. $A_0=-100$
%GeV
%reduces the parameter space by 10 GeV.
From these figures, we find that as $m_{1/2}$ increases BR
 decreases. In the figures, the dependence of the BR on $A_0$ is also shown.
 The relative sizes of the chargino and neutralino diagram contributions change with $A_0$.
  The BR, in most of the parameter space, is around
$10^{-12}$. Choosing a smaller $\kappa_2\over
 \kappa_1$ ratio dcreases the BR ratio further. But this ratio has a lower
bound from the fit to the solar mixing angle. The lowest possible BR is  a
factor of 2 smaller than what is shown in the figure. The BR also can be
reduced by a factor of 4 if we increase the seesaw scale from $3\times
10^{13}$ to $10^{15}$

Turning to the BR($\tau\rightarrow\mu(e)+\gamma$), it is at least 3 order
magnitude below  the current experimental result in the parameter space
allowed by the BR($\mu\rightarrow e+\gamma$).

 In Figs 5 and 6, we show the BRs. for $\mu\rightarrow e+\gamma$ for
$\tan\beta$=30 and 40 for $A_0$ =0 and $A_0=200$ GeV
 in model II.  The $m_0$ is chosen such so that $\Omega h^2$ constraint is
satisfied.
 The $\tau\rightarrow\mu(e)+\gamma$ BR is at least 3-4 order magnitude
below the current experimental limits.

 Furthermore, the parameter space where the BR($\mu\rightarrow
e\gamma$) is
 larger, the other observables e.g. BR($B_s\rightarrow \mu\mu$) can be
observed in the current RUN at the Tevatron. Similarly, for these values,
 $\sigma_{\tilde\chi^0_1-p}$ are also large \cite{adkt} and can
therefore be observable in planned dark matter experiments.
Thus, once the parameter space is narrowed from these different
measurements, this model can be deciphered by the
$\mu\rightarrow e+\gamma$ experiment.

\begin{figure}\vspace{0cm}
    \begin{center}
    \leavevmode
    \epsfysize=8.0cm
    \epsffile[75 160 575 630]{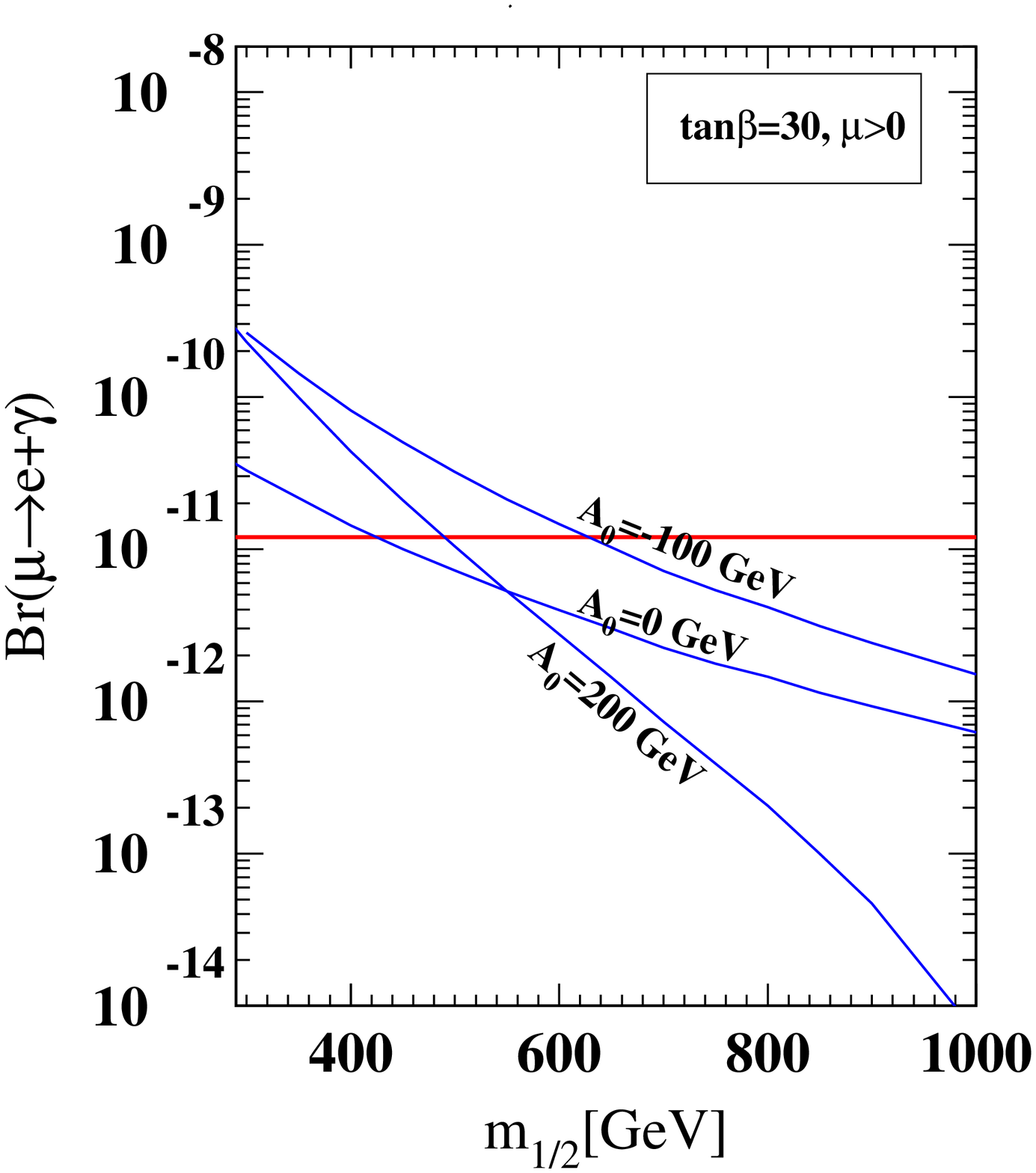}
    %\epsffile[75 160 575 630]{/home/duttabh/lfv/muegmh30mod2.eps}
    \vspace{2.0cm}
     \caption{\label{fig:fig5} BR[$\mu\rightarrow e+\gamma$] as afunction
of $m_{1/2}$ for different values of  $A_0$ for model II.}
\end{center}\end{figure}

\begin{figure}\vspace{0cm}
    \begin{center}
    \leavevmode
    \epsfysize=8.0cm
    \epsffile[75 160 575 630]{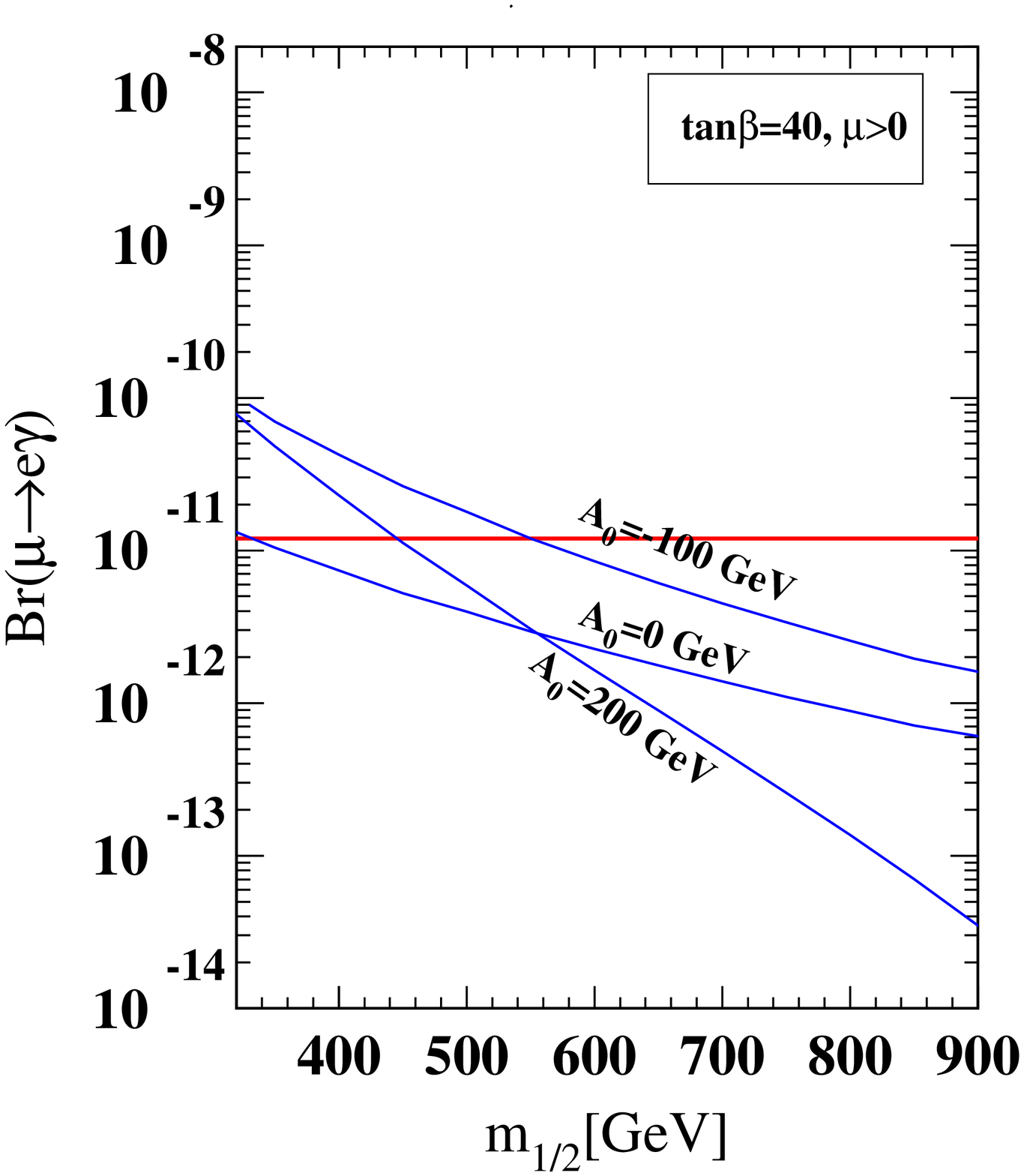}
    %\epsffile[75 160 575 630]{/home/duttabh/lfv/muegmh40mod2.eps}
    \vspace{2.0cm}
     \caption{\label{fig:fig6}  BR[$\mu\rightarrow e+\gamma$] as afunction
of $m_{1/2}$ for different values of  $A_0$ for model II.}
\end{center}\end{figure}

\section{Conclusion}
In summary, we have analysed two models with local SU(2) horizontal
symmetry that lead naturally to bi-large mixing pattern for neutrinos. We
determined the parameters of these models from the detailed numerical fit to the
neutrino masses and the mixing angles. Using these models we then
obtained  predictions for lepton flavor violating
decays BR[$\mu\rightarrow e+\gamma$] in mSUGRA framework. The parmeter space, we
consider, is constrained by the following experimental
constraints: BR[$b\rightarrow s\gamma$], Higgs mass and the recent relic density
 results from WMAP and the Brookheven g-2 experiment.  We find that, in both models, the
 branching ratio is large in a large region of parameter space and is in
the range accessible to the current round of
searches for this process. This parameter space can simultaneously be explored
at Tevatron, LHC and the upcoming of dark matter detectors.

The work of R. N. M. is supported by the National Science Foundation
 Grant No. PHY-0099544 and that of B. D. by the Natural Sciences and
Engineering Research Council of Canada.


\newpage

 \begin{thebibliography}{99}
\bibitem{review} For recent reviews, see S. Bilenky, C. Giunti, J. Grifols
and E. Masso, hep-ph/0211462 (Phys. Rep., to appear); S. Pakvasa and
J. W. F. Valle, hep-ph/0301061.

\bibitem{kuchi} R. Kuchimanchi and R. N. Mohapatra, Phys. Rev. {\bf D 66},
051301 (2002); Phys. Lett. {\bf B 552}, 198 (2003).

\bibitem{other}   D. S. Shaw and R. R. Volkas, Phys. Rev. {\bf D47},
241 (1993); R. Barbieri, L. Hall and A. Romanino, Phys. Lett. {\bf
B401}, 47 (1999);  K. S. Babu and R. N. Mohapatra, Phys. Rev. Lett.
 {\bf 83}, 2522 (1999); T. Blazek, S. Raby and K. Tobe, Phys. Rev. {\bf
D62}, 055001 (2000); S. Raby, hep-ph/0302027; N. Maekawa, hep-ph/0212141.



\bibitem{seesaw} M. Gell-Mann, P. Rammond and R. Slansky, in {\it
Supergravity}, eds. D. Freedman {\it et al.} (North-Holland, Amsterdam,
1980); T. Yanagida, in proc. KEK workshop, 1979
(unpublished); R.N. Mohapatra and G. Senjanovi\'c,
Phys. Rev. Lett. {\bf 44}, 912 (1980); S. L. Glashow, {\it Cargese
lectures}, (1979).


\bibitem{emutau} R. Barbieri, L. Hall, D. Smith, A. Strumia and
N. Weiner, hep-ph/9807235; A. Joshipura and S. Rindani, Eur.Phys.J. {\bf
C14}, 85 (2000); R. N. Mohapatra, A. Perez-Lorenzana, C. A. de S. Pires,
Phys. Lett. {\bf B474}, 355 (2000);
T. Kitabayashi and M. Yasue,
Phys. Rev. {\bf D 63}, 095002 (2001); Phys. Lett. {\bf B 508}, 85 (2001);
hep-ph/0110303;  L. Lavoura, Phys. Rev. D 62, 093011 (2000);
 W. Grimus and L. Lavoura, Phys. Rev. D 62, 093012 (2000);
 J. High Energy Phys. 09, 007 (2000); J. High Energy Phys. 07, 045 (2001);
R. N. Mohapatra, hep-ph/ 0107274; Phys. Rev. {\bf D 64}, 091301 (2001);
 K. S. Babu and R. N. Mohapatra, Phys. Lett. {\bf B 532}, 77
(2002); H. S. Goh, R. N. Mohapatra and S.-P. Ng,
hep-ph/0205131; Phys. Lett. {\bf B 542}, 116 (2002); Duane A. Dicus,
Hong-Jian He, John N. Ng, Phys. Lett. {\bf B 536}, 83 (2002); Q. Shafi and
Z. Tavartkiladze, Phys. Lett. {\bf B 482}, 1451 (2000); Mass matrices
with $L_e-L_{\mu}-L_{\tau}$ symmetry were
discussed in S. Petcov, Phys. Lett. {\bf B 110 }, 245 (1982).

\bibitem{lfv1} F. Borzumati and A. Masiero, Phys. Rev. Lett. {\bf 57}, 961
(1986);  L. Hall, V. Kostelecky and  S. Raby, Nucl. Phys. {\bf
B267}, 415, (1986).
\bibitem{lfv2} J. Hisano, T. Moroi, K. Tobe, M. Yamaguchi and T. Yanagida, Phys. Lett. {\bf B357},
 579 (1995);
J. Hisano, T. Moroi, K. Tobe and M. Yamaguchi, Phys. Rev. {\bf D53}, 2442 (1996).
\bibitem{lfv3} S. F. King and M. Oliveira, Phys. Rev. {\bf D60}, 035003 (1999);
J. Hisano and D. Nomura, Phys. Rev. {\bf D59}, 116005
(1999);
K. S. Babu, B. Dutta and R. N. Mohapatra, Phys. Lett. {\bf B458}, 93 (1999);
W. Buchmuller, D. Delepine and F. Vissani Phys. Lett. {\bf B459}, 171 (1999);
J. Ellis, M. Gomez, G. Leontaris, S. Lola and D. Nanopoulos, Eur. Phys. J. {\bf C14}, 319 (2000);
T. Blazek and S. King, Phys. Lett. {\bf B518}, 109 (2001);
S. Lavgnac, I. Masina and C. Savoy, Phys. Lett. {\bf B520}, 269 (2001);
Nucl. Phys. {\bf B633}, 139 (2002);
J. Casas and A. Ibarra, Nucl. Phys. {\bf B618}, 171 (2001);
R. Gonzalez Felipe and F. Joaquim, JHEP {\bf 0109}, 015 (2001);
J. Sato, K. Tobe and T. Yanagida, Phys. Lett. {\bf B498}, 189 (2001);
X. J. Bi, Y. B. Dai and X. Y. Qi,
Phys. Rev. {\bf D63}, 096008 (2001), Phys. Rev. {\bf D66}, 076006 (2002);
G. Cvetic, C. Dib, C.S. Kim and J.D. Kim,
Phys. Rev. {\bf D66}, 034008 (2002); J. Ellis, J. Hisano, S. Lola and M. Raidal,
Nucl. Phys. {\bf B621}, 208 (2002);
A. Kageyama, S. Kaneko,N. Shimoyama and M. Tanimoto,
Phys. Lett. {\bf B527}, 206 (2002);
D. Chang, A. Masiero and H. Murayama,  hep-ph/0205111;
J. Ellis, J. Hisano, M. Raidal and Y. Shimizu, Phys. Rev. {\bf D66},
115013 (2002);
F. Deppisch, H. Pas, A. Redelbach, R. Ruckl and Y. Shimizu, hep-ph/0206122;
K. Babu and C. Kolda, Phys. Rev. Lett.{\bf 89}, 241802 (2002); A. Rossi,
Phys. Rev. {\bf D66}, 075003 (2002);
I. Masina and C. Savoy, hep-ph/0211283;  A. Masiero, S. Vempati and
O. Vives, Nucl. Phys. {\bf B649}, 189 (2003); M. Raidal and A. Strumia,
Phys. Lett. {\bf B553}, 72 (2003); T. Fukuyama, T. Kikuchi and N. Okada,
hep-ph/0304190; S. Kaneko, M. Katsumata and M. Tanimoto, hep-ph/0305014.

\bibitem{bdm} K. S. Babu, B. Dutta and R. N. Mohapatra,
 Phys. Rev. {\bf D 67}, 076006 (2003).

\bibitem{psi} See meg.web.psi.ch/docs/progress/jun2002/report.ps.

%\bibitem{bahcall} V. Barger, K. Whisnant, D. Marfatia and
%B. P. Wood,
%hep-ph/0204253; J. Bahcall, C. Gonzalez-Garcia and C. Pena-Garay,
%hep-ph/0204314; A. Bandopadhyay, S. Choubey, S. Goswami and D. P. Roy,
%hep-ph/0204286; P. de Holanda and A. Smirnov, hep-ph/0205241; M. Maltoni,
%T. Schwetz, M. Tortola and J. W. F. Valle, hep-ph/0207227.

%\bibitem{numi} M. Szeleper and A. Para, hep-ex/0110032; M. Diwan et
%al. BNL-69395 (2002).

%\bibitem{JHF} Y. Itow et al. (JHF Collaboration), hep-ex/0106019.
\bibitem{wmap}C. L. Bennett et al, astro-ph/0302207;  D. N. Spergel et
al, astro-ph/0302209.

\bibitem{sugra}A. Chamsheddine, R. Arnowitt and P. Nath, {\it N=1
Supergravity}, World Scintific, Singapore (1984);
R. Barbieri, S. Farrara and C. Savoy, Phys. Lett. {\bf B119}, 343 (1982);
 L. Hall, J. Lykken and S. Weinberg, Phys. Rev.{\bf D27}, 2359 (1983).

\bibitem{kribs} G. Kribs, hep-ph/0304256.

\bibitem{witten} E. Witten, Phys. Lett. {\bf B117}, 324 (1982).

\bibitem{fram} For application of the $3\times 2$ seesaw to leptogenesis,
see P. Frampton, S. L. Glashow and T. Yanagida, Phys. Lett. {\bf B 548},
119 (2002); T. Endoh, S. Kaneko, S. Kang, T. Morozumi and M. Tanimoto,
Phys. Rev. Lett. {\bf 89}, 231601 (2002).
\bibitem{fl}G. Fogli, G. Lettera, E. Lisi, A. Marrone, A. Palazzo and A. Rotunno,
 Phys. Rev. {\bf D66}, 093008 (2002). 

%\bibitem{susy} H. Baer and C. Balasz, hep-ph/0303114; H. Baer, C. Balazs,
%A. Belyaev, J. Mizukoshi, X. Tata, Y. Wang, hep-ph/0210441;

\bibitem{coan}J. Ellis, T. Falk and  K. Olive, Phys. Lett. {\bf B444}, 367 (1998);
J. Ellis, T. Falk, K. Olive and M. Srednicki, Astropart. Phys. {\bf 13},
{181} {(2000)}; R. Arnowitt, B. Dutta and Y. Santoso, hep-ph/0010244; Nucl.
Phys. {\bf B606}, {59} {(2001)}; J Ellis, T. Falk, G. Ganis, K. Olive and
M. Srednicki, Phys. Lett. {\bf B570}, {236} {(2001)}; Erratum-ibid. {\bf
15}, 413 (2001); M. Gomez and J. Vergados, hep-ph/0012020; M. Gomez, G.
Lazarides and C. Pallis, Phys. Rev. {\bf D61}, {123512} {(2000)}; Phys.
Lett. {\bf B487}, {313} (2000).
\bibitem{b35} L. Ibanez and C. Lopez, Nucl. Phys {\bf B 233}, {511} {(1984)}.
\bibitem{wmap2} J. Ellis, K. Olive, Y. Santoso, V. Spanos, hep-ph/0303043; H. Baer and C. Balasz, hep-ph/0303114;
R. Arnowitt. B. Dutta, T. Kamon, V. Khotilovich, talk presented at SUGRA20.
\bibitem{ads} R. Arnowitt, B. Dutta, B. Hu and Y. Santoso, Phys. Lett. {\bf B505}, 177 (2001).
\bibitem{adkt} R. Arnowitt, B. Dutta, T. Kamon and M. Tanaka, Phys. Lett. {\bf B538}, 121 (2002).
\end{thebibliography}
\end{document}